\documentclass[a4paper,11pt,english,oneside]{article}
\usepackage{geometry}
\usepackage{amsmath}
\usepackage{amssymb}
\usepackage{color}
\usepackage{graphicx}
\usepackage{subfig}
\usepackage{booktabs}
\usepackage{verbatim}
\usepackage{float}
\usepackage[]{units}
\usepackage{enumitem}
\usepackage[mathcal]{euscript}
\DeclareMathAlphabet{\mathpzc}{OT1}{pzc}{m}{it}
\usepackage[T1]{fontenc}
\usepackage[latin9]{inputenc}
\usepackage{float}
\usepackage{babel}

\providecommand{\tabularnewline}{\\}
\newcommand{\hub}{\tilde{h}_u}
\newcommand{\hdb}{\tilde{h}_d}
\newcommand{\hu}{h_u}
\newcommand{\hd}{h_d}

\newcommand{\br}[1]{\left( #1 \right)}
\newcommand{\abs}[1]{\left| #1 \right|}

\newcommand{\TeV}{\,\mathrm{TeV}}
\newcommand{\GeV}{\,\mathrm{GeV}}

\newcommand{\mgut}{{M_\text{GUT}}}

\newcommand{\ord}[1]{\mathcal{O}\left( #1 \right)}

\newcommand{\eq}[1]{eq.~(\ref{eq:#1})}

\newcommand{\nohyphens}%
{\hyphenpenalty=10000\exhyphenpenalty=10000\relax}

\newcommand{\GSM}{G_\text{SM}}

\newcommand{\U}{\mathcal{U}}
\newcommand{\SU}{\mathcal{SU}}
\newcommand{\gh}[1]{\mathpzc g_{#1}}

\allowdisplaybreaks[1]
\usepackage{hyperref}

\newlength{\myem}
\newcounter{mysubequation}[equation]

\newcommand{\SISSA}{SISSA/ISAS and INFN, I--34151 Trieste, Italy}

\newcommand{\preprintnumber}{%
SISSA--xx/20yy/EP}

\newcommand{\titletext}{Viable and simplified semi-direct gauge mediation with the 4--1 model} 
\newcommand{\authortext}{\large Francesco Caracciolo
\medskip\\\em\normalsize 
\SISSA
}

\newcommand{\abstracttext}{We present a simple and phenomenologically acceptable extension of the 4--1 model of dynamical supersymmetry breaking, 
in which messengers and MSSM superfields are directly coupled to the hidden sector without participating in the supersymmetry breaking mechanism; 
although parametrically suppressed by a loop factor, gaugino masses turn out to be comparable to sfermion masses because of the presence of enhancing 
factors ultimately due to the different origin of the gaugino and the sfermion mass terms. We also describe what can be considered the simplest 
realization of the Higgs sector and how electroweak symmetry breaking can take place in this model.
Finally, in the Appendix, we have listed a set of closed-form expressions for the computation of $A$--terms and soft squared masses of light 
scalars at the messenger scale in the presence of the most general form of "matter-messenger" couplings, extending in this way 
some results already known from the literature.}

\title{
\normalsize
\hspace*{\fill}
\begin{tabular}[t]{l}\preprintnumber\end{tabular}
\vspace{3\baselineskip}\\\Large
\bfseries\titletext\bigskip}

\author{\begin{minipage}[t]{0.8\textwidth}
\normalsize\centering\authortext
\end{minipage}}
\date{}

\begin{document}
\bigskip
\maketitle
\begin{abstract}\normalsize\noindent
\abstracttext
\end{abstract}\normalsize\vspace{\baselineskip}
\section{Introduction}

Many models with gauge-mediated dynamically broken supersymmetry run into phenomenological problems, among which the most 
serious is represented by the gaugino screening \cite{ArkaniHamed:1998kj}. It has been shown in \cite{Caracciolo:2012de} 
that it is possible to overcome this difficulty if the sources of the soft mass terms for sfermions and gauginos are 
different and, in particular, if sfermions are charged under an extra gauge symmetry which is broken at a very high scale 
and gauginos acquire a soft Majorana mass term\footnote{The possibility of Dirac masses has been studied 
in \cite{Fayet:1978qc}} through the coupling of the messengers to the hidden sector. This mechanism of mediation of 
supersymmetry breaking effects, providing a dynamical realization of tree-level gauge mediation \cite{Nardecchia:2009ew}, 
can be seen as a simplified and viable version of the semi-direct gauge mediation framework \cite{Seiberg:2008qj}, 
where only messengers are directly coupled to the hidden sector.

In this paper, we want to show a simple extension of the 4--1 calculable model of dynamical supersymmetry breaking
 \cite{Dine:1995ag} which is based on the following scheme: the hidden sector is composed of the 4--1 model in which the 
strong $\SU(4)$ dynamics generates, at the non-perturbative level, a term in the superpotential whose main effect lies in 
the stabilization of the vacuum of the theory in a point in which both supersymmetry and gauge symmetry are broken 
spontaneously, making, in particular, the vacuum expectation value of the $\U(1)$ D--term non vanishing; if the observable 
chiral superfields are charged under the $\U(1)$ factor, then a soft mass term is generated for the sfermions.

The outline of the paper is the following: in Section~\ref{sec:4--1} we summarize and review the features of the 4--1 model
 which are necessary for the understanding of the rest of the paper; in Section~\ref{sec:coupl} we describe how to couple 
the observable sector to the hidden sector, exploring the implications in the low-energy spectrum of the theory; in 
Section~\ref{sec:Higgs} we show one possible realization of the Higgs sector and comment about the breaking of the 
electroweak symmetry; in Appendix A we explain how it is possible to recover the MSSM at low energies and in 
Appendix B we list a set of closed-form expressions for the computation of $A$--terms and soft-squared mass terms for 
light fields in the most general setup of gauge mediation which includes matter-messenger couplings, 
extending some of the results already known in the literature \cite{Craig:2012xp,Chacko:2001km}.

\section{The 4--1 sector}
\label{sec:4--1}

We first review the 4--1 model, which is based on the $\SU(4)\times\U(1)$ gauge group and matter content as in 
Table \ref{tab:41}.

For later convenience, we collect the $\chi$'s in a matrix $A$:
\begin{equation}
A=\left(
\begin{matrix}
0 && \chi_1 && \chi_2 && \chi_4 \\
-\chi_1 && 0 && \chi_3 && \chi_5 \\
-\chi_2 && -\chi_3 && 0 && \chi_6 \\
-\chi_4 && -\chi_5 && -\chi_6 && 0 
\end{matrix}\right).
\end{equation}
If we assume perturbativity of the $\U(1)$ dynamics at the scale $\Lambda$ at which the $\SU(4)$ gauge 
coupling $\gh{4}$ becomes strong, then the  superpotential can be written as the sum of a classical and a 
non-perturbative term:
\begin{equation}
W_{\text{4--1}}=W_\text{cl}+W_\text{np}
\end{equation}
where
\begin{equation}
W_\text{cl}=hS\overline{F}_i F_i
\end{equation}
\begin{table}[H]
\begin{centering}
\begin{tabular}{ccc}
 & $\SU(4)$ & $\U(1)$\tabularnewline
$S$ & $1$ & $4$\tabularnewline
$F$ & $4$ & $-3$\tabularnewline
$\overline{F}$ & $\overline{4}$ & $-1$\tabularnewline
$\chi$ & $6$ & $2$\tabularnewline
\end{tabular} .
\par\end{centering} 
\caption{matter content of the 4--1 model.}
\label{tab:41}
\end{table}
\noindent
and
\begin{equation}
W_\text{np}=2\dfrac{\Lambda^5}{\sqrt{\overline{F}_i F_j A_{ik} A_{lm} \epsilon^{jklm}} },
\end{equation}
where $\epsilon^{jklm}$ is the totally antisymmetric tensor with $\epsilon^{1234}=1$.

If we assume a hierarchy of the coupling constants $\gh{4} \gg \gh{1} \gg h$ (where $\gh{1}$ is the $\U(1)$ gauge 
coupling constant) and $h\ll 1$, then the vacuum of the theory is calculable and can be found as a power expansion
 in the parameters $h/\gh{4}$ and $h/\gh{1}$. In particular, at the lowest order, the vacuum is located at
\begin{equation}
A=\left(\begin{array}{cccc}
0 & \frac{a}{\sqrt{2}} & 0 & 0 \\
-\frac{a}{\sqrt{2}} & 0 & 0 & 0 \\
0 & 0 & 0 & \frac{a}{\sqrt{2}} \\
0 & 0 & -\frac{a}{\sqrt{2}} & 0 
\end{array}\right)M
\qquad
F=\overline{F}=\left(\begin{array}{c}
b \\
0 \\
0 \\
0
\end{array}\right) M
\qquad
S=c M,
\end{equation}
where
\begin{equation}
M\equiv\frac{\Lambda}{h^{\frac{1}{5}}}\gg \Lambda
\end{equation}
\begin{equation}
c\equiv\sqrt{b^2 - \frac{a^2}{2}}
\end{equation}
and, approximately, $a\approx 1.492$, $b\approx 1.102$. At the lowest order, furthermore, the non-zero vacuum 
expectation values of the F--terms are the following:
\begin{equation}
F_{\chi_1}=F_{\chi_6}=-\frac{\sqrt{2}}{a^2 b}F \qquad F_{F_1}=F_{\overline{F}_1}=\frac{a b^3 c-1}{a b^2}F \qquad F_S = b^2 F,
\end{equation}
where
\begin{equation}
F=h^{\frac{3}{5}}\Lambda^2.
\end{equation}
We are now in position to infer that, in the vacuum, D--terms are non-zero and, in particular, the $\U(1)$ D--term 
is\footnote{A convenient way to gain information on the vacuum structure of the model is the following: in the vacuum, 
F--terms, D--terms and scalars $\phi$ are not independent: $F^\dagger T_a F = \sum_b g_b^2 D_b \phi^\dagger T_b T_a \phi$, 
where $T_a$ is any symmetry global/gauge symmetry generator, the index $b$ labels the gauge symmetry generators and $g_b$ 
is the gauge coupling constant of the group which the generator $T_b$ belongs to; it is easy to see that once the 
zeroth-order vacuum expectation values of $\phi$ and F--terms in the expansion in $h/\gh{1}$ and $h/\gh{4}$ are known, 
then the aforementioned relation yields the vacuum expectation values of the D--terms at the second-order in $h/\gh{1}$ 
and $h/\gh{4}$.}
\begin{equation}
D_{\U(1)} = \dfrac{4b^2 - 2a^2 + a^6 b^6 + 2a^3 b^3 \sqrt{4b^2 - 2a^2}}{2a^4 b^4 (6b^2 - a^2)} \frac{1}{\gh{1}^2}
\frac{F^2}{M^2}\equiv d \frac{1}{\gh{1}^2} \frac{F^2}{M^2},
\end{equation}
where $d\approx 0.349$.

We can now couple the 4--1 model to the MSSM and messenger sectors.

\section{Coupling the observable sector to the 4--1 model}
\label{sec:coupl}

The aim of this section is to show an extension of the 4--1 model, which from now on we will identify with the hidden 
sector, based on the gauge group $\SU(4)\times\U(1) \times\GSM$, in which the scalar components of the would-be MSSM 
superfields, charged under the $\U(1)$ gauge group, gain a positive soft squared mass via the $\U(1)$ D--term coupling 
and in which it is possible to identify a set of messenger superfields which give mass to gauginos at the one-loop level, 
as in the standard gauge mediation framework \cite{Dine:1981za}. 

Let us consider the spectrum in Table \ref{tab:41coupl}, where $j$ is a family index which runs from $1$ to $3$, $i$, 
which can take values $1$, $2$,..., $n_m$, labels the messengers $\varphi$, $\tilde{\varphi}$, $\psi$ and $\tilde{\psi}$ 
within each family; moreover, for reasons that soon will become apparent, we require that one among the following $\U(1)$
 charge 
assignments is realized: either
\begin{align}
q_m & =-2-\sqrt{9y^2-12n_m y+\frac{4}{3}\br{4n_m^2-1}}\nonumber \\
\tilde{q}_m & =-2+\sqrt{9y^2-12n_m y+\frac{4}{3}\br{4n_m^2-1}}
\label{eq:pos1}
\end{align}
or
\begin{align}
q_m & =-2+\sqrt{9y^2-12n_m y+\frac{4}{3}\br{4n_m^2-1}} \nonumber \\
\tilde{q}_m & =-2-\sqrt{9y^2-12n_m y+\frac{4}{3}\br{4n_m^2-1}};
\label{eq:pos2}
\end{align}
the superfields $q$, $l$, $\tilde{u}$, $\tilde{d}$, $\tilde{e}$ and $\tilde{n}$ are the light observable fields; 
the superpotential is:
\begin{align}
& W=W_{\text{4--1}}+W_{\text{mess}}\nonumber\\
& W_{\text{mess}}=\lambda^1_{ij,kl}S\varphi_i^k\tilde{\varphi}_j^l+\lambda^2_{ij,kl}S\psi_i^k\tilde{\psi}_j^l.
\label{eq:mess}
\end{align}
The $\U(1)$ charge assignment described in Table~\ref{tab:41coupl} (with one among \eq{pos1} and \eq{pos2} satisfied) is 
derived imposing the following conditions: family independence of the $\U(1)$ charges; all of the 
$\varphi$'s ($\tilde{\varphi}$'s) and all of the $\psi$'s ($\tilde{\psi}$'s) have the same $\U(1)$ charges; cancellation of 
all the anomalies; the superpotential in \eq{mess} is allowed; the requirement that the charges of $l$, $\tilde{e}$, $q$ 
and $\tilde{d}$ (denoted with $q_l$, $q_{\tilde{e}}$, $q_q$ and $q_{\tilde{d}}$ respectively) are not independent: 
$q_l+q_{\tilde{e}}=q_q+q_{\tilde{d}}$ (the reason for this requirement lies in the coupling to the Higgs).
\begin{table}[H]
\begin{centering}
\begin{tabular}{ccc|c}
& $\SU(4)$ & $\U(1)$ & $\GSM$\tabularnewline
$S$ & $1$ & $4$ & 1\tabularnewline
$F$ & $4$ & $-3$ & 1\tabularnewline
$\overline{F}$ & $\overline{4}$ & $-1$ & 1\tabularnewline
$\chi$ & $6$ & $2$ & 1\tabularnewline
\hline 
$\varphi_j^i$ & 1 & $q_m$ & $\left(3,1,-1/3 \right) $ \tabularnewline
$\tilde{\varphi}_j^i$ & 1 & $\tilde{q}_m$ & $ \left(\overline{3},1,1/3\right)$ \tabularnewline
$\psi_j^i$ & 1 & $q_m$ & $ \left(1,\overline{2},1/2 \right) $ \tabularnewline
$\tilde{\psi}_j^i$ & 1 & $\tilde{q}_m$ & $\left(1,2,-1/2 \right) $ \tabularnewline
\hline
$q_j$ & 1 & $y$ & $\left(3,2,1/6 \right)$ \tabularnewline
$l_j$ & 1 & $4n_m-3y$ & $\left(1,2,-1/2 \right)$ \tabularnewline
$\tilde{u}_j$ & 1 & $y$ & $\left(\overline{3},1,-2/3 \right)$ \tabularnewline
$\tilde{d}_j$ & 1 & $4n_m-3y$ & $ \left(\overline{3},1,1/3 \right)$ \tabularnewline
$\tilde{e}_j$ & 1 & $y$ & $\left(1,1,1 \right)$ \tabularnewline
$\tilde{n}_j$ & 1 & $5y$ & $\left(1,1,0 \right) $\tabularnewline
\end{tabular}
\par\end{centering}
\caption{spectrum of the theory (without the Higgs sector; the $\GSM$ representations of the messengers 
$\phi$, $\tilde{\phi}$, $\psi$ and $\tilde{\psi}$ are chosen in such a way that it is possible to embed 
$\big( \phi,\psi \big)$ and $\big( \tilde{\phi},\tilde{\psi}\big)$ in complete $SU(5)_\text{GUT}$ multiplets; 
$q$, $l$, $\tilde{u}$, $\tilde{d}$, $\tilde{e}$ and $\tilde{n}$ denote the MSSM superfield together with a right-handed 
neutrino $\tilde{n}$.}
\label{tab:41coupl}
\end{table}
\noindent
We finally require that $q_m$ and $\tilde{q}_m$ (as well as $y$) are rational \cite{Banks:2010zn}.

The spectrum so defined does not destabilize the vacuum of the 4--1 model because messengers have a large supersymmetric 
mass term via the coupling in \eq{mess}, while light sfermions acquire a \textit{positive} soft squared mass term via 
the $\U(1)$ D--term coupling provided $0<y<4/3n_m$:
\begin{equation}
\tilde{m}^2_{q,\tilde{u},\tilde{e}}=yd\frac{F^2}{M^2} \qquad \tilde{m}^2_{l,\tilde{d}}=\br{4n_m-3y}d\frac{F^2}{M^2}\qquad
\tilde{m}^2_{\tilde{n}}=5yd\frac{F^2}{M^2}.
\label{eq:scalarmasses}
\end{equation}
Gaugino masses arise at the one-loop level; at the messenger scale
\begin{equation}
M_i (M_\text{mess})= 3 n_m \frac{b^2}{\sqrt{b^2-\frac{a^2}{2}}} \frac{\alpha_i(M_\text{mess})}{4\pi} \frac{F}{M};
\label{eq:gauginos}
\end{equation}
since at the one-loop level the gaugino mass evaluated at a generic scale $\mu$ is given by \eq{gauginos} with the 
substitution $\alpha_i(M_\text{mess})\rightarrow \alpha_i(\mu)$, we find at the $\TeV$ scale the following relation:
\begin{equation}
M_3 \approx 0.13 \frac{n_m}{\sqrt{y}}\tilde{m}_{q},
\label{eq:gluino}
\end{equation}
where $\tilde{m}_{q}$ is defined in \eq{scalarmasses}; the ratio between the gluino and $\tilde{m}_{q}$ depends on the 
free parameters $n_m$ and $y$, which can be chosen in such a way to make  $\tilde{m}_{q}$ not hierarchically larger than 
the gluino mass.

In order to keep perturbativity of the gauge coupling constants up to the unification scale, a lower bound on the 
messenger masses should be imposed and, in the simple case in which messenger masses are almost degenerate, it is possible 
to estimate this value \cite{Giudice:1998bp}:
\begin{equation}
\frac{M_{\text{mess}}}{\mgut} \gtrsim e^{-\frac{50}{n_m}}.
\label{eq:lowbound}
\end{equation}
From \eq{gluino} and \eq{lowbound}, it is possible to deduce that, in general, a reduction of the hierarchy between gluino 
and sfermion masses can be obtained if the lower bound on the messenger mass is raised.
To have an idea of the energy scales and of the parameters involved in the theory let us consider the following 
situation: $n_m=5$, $y=4$;
in this case, it is found that
\begin{equation}
M_3 \approx 0.33 \tilde{m}_{q}
\end{equation}
and that messenger masses have the following bound
\begin{equation}
 M_\text{mess}> 10^{12} \GeV;
\end{equation}
this implies that the Yukawa coupling $h$ that appears in $W_{\text{4--1}}$ is constrained by the following 
relation\footnote{In \eq{h} we have assumed that $M_\text{mess}\approx M$.}:
\begin{equation}
 h=\frac{F}{M^2} =\frac{F}{M}\frac{1}{M} \lesssim 10^{-9}\frac{\tilde{m}}{1\TeV},
\label{eq:h}
\end{equation}
which is compatible with the hypothesis $h\ll 1$.

It should be finally mentioned that it is possible to require the existence of messenger-matter couplings in the 
superpotential, which may be necessary to raise the Higgs boson mass via the generation of sizeable $A$--terms 
(see Appendix B); such a situation can be realized, for example, if we take $y=32/5$ and $n_m=5$; indeed, in this 
case $q_m=-64/5$ and terms which couple $\varphi$ or $\psi$ to $q$, $\tilde{u}$ and $\tilde{e}$ are allowed.
\section{The Higgs sector and the Yukawa interactions}
\label{sec:Higgs}
To complete the model, we have to introduce the Higgs fields and their couplings to the matter; since this is highly 
model-dependent, we present here what we consider the simplest realization of the Higgs sector.

We add to the spectrum in Table~\ref{tab:41coupl} four Higgs superfields in a vectorlike representation of the gauge 
group, see Table~\ref{tab:41couplhiggs}, and, at the renormalizable level, the following superpotential is allowed:
\begin{align}
& W=W_{\text{4--1}}+W_{\text{mess}} +W_{\text{Yukawa}}+W_{\text{Higgs}}\nonumber\\ 
& W_{\text{Yukawa}}= \lambda^u_{ij} h_u q_i \tilde{u}_j + \lambda^d_{ij} h_d q_i \tilde{d}_j + \lambda^e_{ij} h_d l_i 
\tilde{e}_j. 
\label{eq:yukawa}
\end{align}
In writing \eq{yukawa}\footnote{Notice that all the interactions between the observable sector and light fields present 
in the spectrum of the hidden sector are suppressed by factors of $\ord{1/M}$, so that they can be considered completely 
harmless for low-energy phenomenology.} we have required that $h_u$ and $h_d$ either have a component or can be identified 
with the MSSM up and down-type Higgs superfields.
\begin{table}[H]
\begin{centering}
\begin{tabular}{ccc|c}
& $\SU(4)$ & $\U(1)$ & $\GSM$\tabularnewline
$S$ & $1$ & $4$ & 1\tabularnewline
$F$ & $4$ & $-3$ & 1\tabularnewline
$\overline{F}$ & $\overline{4}$ & $-1$ & 1\tabularnewline
$\chi$ & $6$ & $2$ & 1\tabularnewline
\hline 
$\varphi_j^i$ & 1 & $q_m$ & $\left(3,1,-1/3 \right) $ \tabularnewline
$\tilde{\varphi}_j^i$ & 1 & $\tilde{q}_m$ & $ \left(\overline{3},1,1/3\right)$ \tabularnewline
$\psi_j^i$ & 1 & $q_m$ & $ \left(1,\overline{2},1/2 \right) $ \tabularnewline
$\tilde{\psi}_j^i$ & 1 & $\tilde{q}_m$ & $\left(1,2,-1/2 \right) $ \tabularnewline
\hline
$q_j$ & 1 & $y$ & $\left(3,2,1/6 \right)$ \tabularnewline
$l_j$ & 1 & $4n_m-3y$ & $\left(1,2,-1/2 \right)$ \tabularnewline
$\tilde{u}_j$ & 1 & $y$ & $\left(\overline{3},1,-2/3 \right)$ \tabularnewline
$\tilde{d}_j$ & 1 & $4n_m-3y$ & $ \left(\overline{3},1,1/3 \right)$ \tabularnewline
$\tilde{e}_j$ & 1 & $y$ & $\left(1,1,1 \right)$ \tabularnewline
$\tilde{n}_j$ & 1 & $5y$ & $\left(1,1,0 \right) $\tabularnewline
\hline
$h_u$ & 1 & $-2y$ &   $\left(1,2,\frac{1}{2} \right)$ \tabularnewline
$h_d$ & 1 & $-4n_m+2y$ &   $\left(1,2,-\frac{1}{2} \right)$ \tabularnewline
$\tilde{h}_u$ & 1 & $2y$ &   $\left(1,2,-\frac{1}{2} \right)$ \tabularnewline
$\tilde{h}_d$ & 1 & $4n_m-2y$ &   $\left(1,2,\frac{1}{2} \right)$ \tabularnewline
\end{tabular}
\par\end{centering}
\caption{spectrum of the theory, with the Higgs sector}
\label{tab:41couplhiggs}
\end{table}
\noindent






We are now in the position to describe two scenarios which lead to a phenomenologically acceptable pattern of electroweak 
symmetry breaking, without attempting neither to solve the $\mu$--problem nor to explain possible hierarchies of the 
parameters in the Higgs sector.

The simplest choice consists in the assumption that the Higgs superpotential is
\begin{equation}
W_{\text{Higgs}} = \mu_u h_u \tilde{h}_u+ \mu_d h_d \tilde{h}_d,
\label{eq:WH1}
\end{equation}
where we consider the situation in which $\mu_u$ and $\mu_d$ are $\ord{\TeV}$ and, as a consequence, in the end we find four Higgs superfields 
at the $\TeV$ scale. Since, at the messenger scale, the soft terms generated in the Higgs sector accidentally conserves the $\U(1)$ symmetry, 
then a $\U(1)$ preserving $B\mu$--term is radiatively generated below the scale $M$; in this context, then, it is possible 
to conclude as in \cite{Caracciolo:2012de}: only one out of the two pairs of doublets $\big( h_u,\tilde{h}_u \big)$ and 
$\big( h_d,\tilde{h}_d \big)$ acquires a non-vanishing vacuum expectation value; in order to have non-vanishing vacuum expectation 
values for $h_u$ and $h_d$, it is necessary to break the accidental $\U(1)$ symmetry and this can be done via small 
non-renormalizable interactions. 
 
Alternatively, it is possible to have two Higgs superfields at the $\TeV$ scale if we consider the following tree-level Higgs superpotential
\begin{equation}
W_{\text{Higgs}} = \mu_u h_u \tilde{h}_u+ \mu_d h_d \tilde{h}_d+W_{\text{non-ren}},
\label{eq:WH2}
\end{equation}
where $W_{\text{non-ren}}$ is the non-renormalizable part of the superpotential, in which terms like 
$\br{F\overline{F}}^n  /  \br{\Lambda^{2n-1}} \tilde{h}_u \tilde{h}_d$ and $S^{n} / \Lambda^{n-1}h_u h_d$ are included, 
generating $\U(1)$-breaking $B\mu$--term. Such a possibility is explored in Appendix A, where it is shown that the MSSM 
with two Higgs doublets can be recovered at low energies.





\section{Summary}
\label{sec:conclusions}

We have proposed a phenomenologically viable model, based on a simple extension of the 4--1 model, in which 
supersymmetry is broken dynamically and communication of the supersymmetry breaking effects in the observable sector 
is achieved via renormalizable non-SM gauge interactions \cite{Nardecchia:2009ew}.

In particular, we have realized a situation in which the tree-level generated soft sfermion masses are flavor independent, 
solving in this way the supersymmetric flavor problem; gaugino masses arise at the one-loop level, as in the minimal 
gauge mediation mechanism; because of the presence of various enhancing factors, the hierarchy between sfermions and 
gauginos is reduced in a remarkable and acceptable way.

In the situation described in the current paper, the Higgs sector is highly model-dependent; we have proposed what we 
think can be considered its simplest realization. In particular, in Appendix A, we explore the possibility to recover at low 
energies the MSSM.

In Appendix B, we list closed-form expressions for $A$--terms and soft squared masses generated at the messenger scale 
in models of gauge mediation with generic "matter-messenger" couplings.

\section{Acknowledgments}

I am grateful to Andrea Romanino for his guide during the preparation of the work and the writing of the manuscript; I am also grateful to Matteo Bertolini for useful discussions; I would finally like to thank the Galileo Galilei Institute for Theoretical Physics, where part of the work was done, for its hospitality.

\appendix

\section*{Appendix A: recovering the MSSM at low energies}
\label{sec:low}

The purpose of this Appendix is to show that there exists the possibility to recover the two-Higgs-doublet MSSM at 
low energies; since the Higgs sector is highly model dependent, in the specific framework that we are going to describe 
we ignore both the fine-tuning of the parameters and the hierarchies of the coupling constants which identify the 
Higgs sector.

Let us first consider the Higgs sector superpotential including non-renormalizable terms
\begin{equation}
W_\text{Higgs}=\mu_u \hu \hub +\mu_d \hd \hdb + X_1 \hub \hdb + X_2 \hu \hd,
\end{equation}
where $X_1$ and $X_2$ can be considered as spurion fields which acquire both a scalar and an $F-$term vacuum 
expectation values  ($ X_1 =\mu_1+F_1 \theta^2$ and $X_2=\mu_2+F_2 \theta^2$), and K\"ahler potential
\begin{align} 
K_\text{Higgs}= & \hu^\dagger \hu +\hd^\dagger \hd + \hub^\dagger \hub+ \hdb^\dagger \hdb+ \nonumber \\ 
& - \theta^2 \overline{\theta}^2 \br{-m_1^2 \hu^\dagger \hu - m_2^2\hd^\dagger \hd +m_1^2 \hub^\dagger \hub +m_2^2 
\hdb^\dagger \hdb}.
\end{align}
For reasons that will become apparent in a moment, we assume that $\tilde{\mu} \equiv \sqrt{\mu_d^2+\mu_1^2}$ is much 
bigger than any other mass scale appearing in the Higgs sector and that $m_1$, $m_2$, $\mu_u$, $\mu_2$, $F_1 / \mu_1$ 
and $\sqrt{F_2}$ are of the same order of magnitude, namely the $\TeV$ scale. This setup is inspired 
by the model we have analyzed in Section~\ref{sec:Higgs}, where $m_1$ and $m_2$ can be identified with the soft masses 
generated via tree-level gauge mediation and $X_1$ and $X_2$ can be identified, for example, with $\br{F\overline{F}}^n  /  \br{\Lambda^{2n-1}} $ 
and $S^{n} / \Lambda^{n-1}$ respectively, where $\Lambda$ is an ultraviolet cutoff.

For later convenience, we define the following parameters
\begin{align}
& X_i=  \mu_i+\phi_i\nonumber\\
& \tan  \alpha =  \frac{\mu_d}{\mu_1}\nonumber\\
&\tilde{\mu} = \sqrt{\mu_1^2 + \mu_d^2}\nonumber\\
& h_h =   \cos \alpha \hub + \sin \alpha \hd \nonumber\\
& h_l =  -\sin \alpha \hub+\cos \alpha \hd,
\end{align}
where $\phi_i$'s are spurion superfields acquiring only F--term vacuum expectation values.
We write again the superpotential and the K\"ahler potential in terms of the new variables:
\begin{align}
W_{\text{Higgs}} = & \mu_u \hu \br{\cos \alpha h_h - \sin \alpha h_l} + \tilde{\mu} h_h \hdb + \mu_2 h_u \br{\sin \alpha h_h + \cos \alpha h_l}+ \nonumber \\
& + \phi_1 \br{\cos \alpha h_h - \sin \alpha h_l}\hdb + \phi_2 h_u \br{\sin \alpha h_h + \cos \alpha h_l}
\end{align}
\begin{align}
K_{\text{Higgs}} = & \hu^\dagger \hu +  \hd^\dagger \hd+ \hub^\dagger \hub+ \hdb^\dagger \hdb+ \nonumber \\
& - \theta^2 \overline{\theta}^2 \left(- m_1^2 \hu^\dagger \hu+m_1^2 \br{\cos\alpha h_h-\sin\alpha h_l}^\dagger 
\br{\cos\alpha h_h-\sin\alpha h_l}+ \right.\nonumber\\
& \left. - m_2^2 \br{\sin\alpha h_h+\cos\alpha h_l}^\dagger \br{\sin\alpha h_h+\cos\alpha h_l}+m_2^2 \hdb^\dagger \hdb\right)
\end{align}
The assumptions that we have made allow us to integrate out the heavy fields $h_h$ and $\hdb$ in a manifest supersymmetric
 way \cite{Brizi:2009nn}. The effective low-energy superpotential and K\"ahler potential, neglecting more irrelevant 
operators, are:
\begin{equation}
W_{\text{Higgs}}^{\text{eff}}= \br{\mu+\phi_\mu} h_u h_l
\end{equation}
\begin{equation}
K_{\text{Higgs}}^{\text{eff}}=  \hu^\dagger \hu+ h_l^\dagger h_l- \tilde{m}_u^2 \theta^2 \overline{\theta}^2 \hu^\dagger 
\hu- \tilde{m}_l^2 \theta^2 \overline{\theta}^2 h_l^\dagger h_l
\end{equation}
where
\begin{align}
& \mu =  -\mu_u \sin \alpha + \mu_2 \cos \alpha \nonumber\\
& \phi_\mu =  \frac{\phi_1}{\tilde{\mu}} \br{\mu_u \sin\alpha \cos\alpha +\mu_2 \sin\alpha^2}+\cos\alpha \phi_2\nonumber\\
& \tilde{m}_u^2=  -m_1^2\nonumber\\
& \tilde{m}_l^2 =  m_1^2 \sin\alpha^2-m_2^2 \cos\alpha^2-\frac{F_1^2}{\tilde{\mu}^2} \sin\alpha^2.
\end{align}
We have then recovered the MSSM at low energies.

\section*{Appendix B: complete and explicit formulae for the radiatively generated $A$--terms and soft squared masses in theories with matter-messenger couplings}
\label{sec:matmess}

The discovery of a $125\GeV$ Higgs-like particle has raised a problem in the MSSM because, in the absence of $A$--terms, 
the Higgs boson mass is bounded to be smaller than $118\GeV$ if sfermions are lighter than $2\TeV$ \cite{Craig:2012xp}; if, 
on the contrary, $A$--terms are introduced the limits on the Higgs boson mass in the MSSM can be relaxed.

In general models of gauge mediation, the presence of matter-messenger couplings has on one side the r\^ole to generate 
$A$--terms at the one-loop order, but, on the other side, may be dangerous because of the introduction of unwanted 
flavor-violating interactions which should be kept under control.

The goal of this section is to collect a set of useful closed-form formulae for the computation of soft squared masses 
and $A$--terms \textit{at the messenger scale} in general models of gauge mediation with \textit{the most generic} form 
of matter-messenger couplings; this section can be seen as an extension and a generalization of the results derived 
and listed in \cite{Craig:2012xp,Chacko:2001km}.

Let us consider a superpotential containing the following terms:
\begin{equation}
W\supset\frac{1}{2}\lambda_{ijk}\phi_i q_j q_k+\frac{1}{2}\tilde{\lambda}_{ijk}q_i \phi_j \phi_k+
\frac{1}{6}\rho_{ijk}q_i q_j q_k
\end{equation}
where $i$, $j$ and $k$ parametrize both gauge and  flavor indices, the messenger superfields, which are assumed to be 
coupled to a spurion field $X$ such that $\langle X\rangle=M+F\theta^2$, are denoted by $\phi$, the fields $q$ represent 
the observable fields and the following symmetry relations are valid:
\begin{align}
\lambda_{ijk}=\lambda_{ikj} \qquad \tilde{\lambda}_{ijk} = \tilde{\lambda}_{ikj};
\end{align}
moreover, the coefficients $\rho_{ijk}$ can be assumed to be completely symmetric in the indices $i$, $j$ and $k$.

For symmetry reasons it is convenient to write the above expression in the following way:
\begin{align}
W\supset & \frac{1}{6} \br{\lambda^1_{ijk}\phi_i q_j q_k+\lambda^2_{ijk}q_i \phi_j q_k+\lambda^3_{ijk}q_i q_j \phi_k }
+\nonumber\\
& +\frac{1}{6} \br{\tilde{\lambda}^1_{ijk}q_i \phi_j \phi_k+\tilde{\lambda}^2_{ijk}\phi_i q_j \phi_k+
\tilde{\lambda}^3_{ijk}\phi_i \phi_j q_k }+\nonumber\\
&+\frac{1}{6}\rho_{ijk}q_i q_j q_k,
\label{eq:sup1}
\end{align}
where we assume the validity of the following relations:
\begin{align}
\lambda^1_{ijk} = \lambda^2_{jik}=\lambda^3_{jki}=\lambda_{ijk} \qquad \tilde{\lambda}^1_{ijk} = \tilde{\lambda}^2_{jik}=
\tilde{\lambda}^3_{jki}=\tilde{\lambda}_{ijk}
\end{align}
After the integration of the messenger fields we are left with an effective theory in which soft supersymmetry breaking 
interactions appear, in particular:
\begin{equation}
V_\text{soft} \supset \br{\tilde{m}^2}^a_b \tilde{q}_a^\dagger \tilde{q}_b+
\br{ \frac{1}{6}A_{abc}\tilde{q}_a \tilde{q}_b \tilde{q}_c+\text{h.c.}}
\label{eq:softsusy}
\end{equation}
For later convenience we write explicitly the anomalous dimensions of the fields, where the underscripts $+$ and $-$ 
refer respectively to the theory above and below the messenger scale:
\begin{align}
& \gamma_{+q_b}^{q_a}=\frac{1}{16\pi^2}
\br{\frac{1}{2}\rho_{ajk}\rho_{bjk}^{*}+\frac{1}{2}\tilde{\lambda}_{ajk}^{1}\tilde{\lambda}_{bjk}^{1*}+\frac{1}{2}
\lambda_{ajk}^{2}\lambda_{bjk}^{2*}+\frac{1}{2}\lambda^{3}_{ajk}\lambda^{3*}_{bjk}-2C_c(a)g_c^2 \delta_b^a}\nonumber\\
& \gamma_{+\phi_b}^{\phi_a}=\frac{1}{16\pi^2}
\br{\frac{1}{2}\lambda_{ajk}^{1}\lambda_{bjk}^{1*}+
\frac{1}{2}\tilde{\lambda}_{ajk}^{2}\tilde{\lambda}_{bjk}^{2*}+\frac{1}{2}\tilde{\lambda}^{3}_{ajk}
\tilde{\lambda}^{3*}_{bjk}-2C_c(a)g_c^2 \delta_b^a}\nonumber\\
& \gamma_{+q_b}^{\phi_a}=\frac{1}{16\pi^2}\br{\frac{1}{2}\lambda_{ajk}^{1}\rho_{bjk}^{*}+\frac{1}{2}
\tilde{\lambda}_{ajk}^{2}\lambda_{bjk}^{3*}+\frac{1}{2}\tilde{\lambda}^{3}_{ajk}\lambda^{2*}_{bjk}}\nonumber\\
& \gamma_{-q_b}^{q_a}=\frac{1}{16\pi^2}\br{\frac{1}{2}\rho_{ajk}\rho_{bjk}^{*}-2C_c(a)g_c^2 \delta_b^a}\nonumber\\
&\gamma_{-\phi_b}^{\phi_a}=\gamma_{-q_b}^{\phi_a}=0 \qquad \gamma_{+\phi_b}^{q_a}=\gamma_{+q_a}^{\phi_b *}\nonumber\\
& \Delta \gamma_{\psi_a}^{\psi_b}\equiv\gamma_{+\psi_a}^{\psi_b}-\gamma_{-\psi_a}^{\psi_b},
\label{eq:gamma}
\end{align}
where $C_c (a)$ is the quadratic Casimir of the field $q_a$ or $\phi_a$ for the gauge group factor 
with coupling constant $g_c$ and $\psi$ denotes a generic messenger or observable field.

We are now in the position to write down both the $A$--terms and the soft squared masses in \eq{softsusy} induced 
for the fields $q$ for the superpotential in \eq{sup1}:
\begin{equation}
\frac{A_{abc}}{F/M}=-\br{\rho_{dbc}\Delta\gamma^{q_d}_{q_a}+\rho_{adc}\Delta\gamma^{q_d}_{q_b}+\rho_{abd}\Delta\gamma^{q_d}_{q_c}}.
\label{eq:aterms}
\end{equation}
For what concerns the soft squared masses, it is convenient to write down the following expression:
\begin{equation}
\frac{\tilde{m}^2}{\abs{{F/M}}^2} \equiv \tilde{m}^2_1-\tilde{m}^2_2-\tilde{m}^2_3
\end{equation}
where
\begin{align}
\left(\tilde{m}_1^2\right)^{q_i}_{q_j} =  \frac{1}{64\pi^2}  & \left[  \lambda^{2*}_{jbc} \left(\lambda^2_{lbc}
\gamma_{+q_i}^{q_l}+\lambda^2_{ilc}\gamma_{+\phi_b}^{\phi_l}+\lambda^2_{ibl}\gamma_{+q_c}^{q_l}+\tilde{\lambda}^{3}_{lbc}
\gamma_{+q_i}^{\phi_l}+\rho_{ilc}\gamma_{+\phi_b}^{q_l}+\tilde{\lambda}_{ibl}^{1}\gamma_{+q_c}^{\phi_l}\right) + 
\right. \nonumber\\
+ & \lambda^{2}_{ibc} \left(\lambda^2_{lbc}\gamma_{+q_j}^{q_l}+\lambda^2_{jlc}\gamma_{+\phi_b}^{\phi_l}+
\lambda^2_{jbl}\gamma_{+q_c}^{q_l}+\tilde{\lambda}^{3}_{lbc}\gamma_{+q_j}^{\phi_l}+\rho_{jlc}\gamma_{+\phi_b}^{q_l}+
\tilde{\lambda}_{jbl}^{1}\gamma_{+q_c}^{\phi_l}\right)^{*} + \nonumber\\
+ & \lambda^{3*}_{jbc} \left(\lambda^3_{lbc}\gamma_{+q_i}^{q_l}+\lambda^3_{ilc}\gamma_{+q_b}^{q_l}+
\lambda^3_{ibl}\gamma_{+\phi_c}^{\phi_l}+\tilde{\lambda}^{2}_{lbc}\gamma_{+q_i}^{\phi_l}+
\tilde{\lambda}^{1}_{ilc}\gamma_{+q_b}^{\phi_l}+\rho_{ibl}\gamma_{+\phi_c}^{q_l}\right) + \nonumber\\
+ & \lambda^{3}_{ibc} \left(\lambda^3_{lbc}\gamma_{+q_j}^{q_l}+\lambda^3_{jlc}\gamma_{+q_b}^{q_l}+
\lambda^3_{jbl}\gamma_{+\phi_c}^{\phi_l}+\tilde{\lambda}^{2}_{lbc}\gamma_{+q_j}^{\phi_l}+\tilde{\lambda}^{1}_{jlc}
\gamma_{+q_b}^{\phi_l}+\rho_{jbl}\gamma_{+\phi_c}^{q_l}\right)^{*} + \nonumber\\
+ & \tilde{\lambda}^{1*}_{jbc}\left(\tilde{\lambda}^{1}_{lbc}\gamma_{+q_i}^{q_l} +\tilde{\lambda}^{1}_{ilc}
\gamma_{+\phi_b}^{\phi_l} +\tilde{\lambda}^{1}_{ibl}\gamma_{+\phi_c}^{\phi_l} + \lambda^{3}_{ilc}\gamma_{+\phi_b}^{q_l}
+\lambda^{2}_{ibl}\gamma_{+\phi_c}^{q_l}\right)+ \nonumber\\
& \left. +\tilde{\lambda}^{1}_{ibc}\left(\tilde{\lambda}^{1}_{lbc}\gamma_{+q_j}^{q_l}
 +\tilde{\lambda}^{1}_{jlc}\gamma_{+\phi_b}^{\phi_l} +\tilde{\lambda}^{1}_{jbl}\gamma_{+\phi_c}^{\phi_l} 
+ \lambda^{3}_{jlc}\gamma_{+\phi_b}^{q_l}+\lambda^{2}_{jbl}\gamma_{+\phi_c}^{q_l}\right)^{*} \right]
\end{align}

\begin{align}
\left(\tilde{m}_2^2\right)^{q_i}_{q_j} & =  \frac{1}{64\pi^2}  \left[ \rho_{jbc}^{*}
\left( \rho_{lbc}\Delta \gamma_{q_i}^{q_l}+\rho_{ilc}\Delta \gamma_{q_b}^{q_l}
+\rho_{ibl}\Delta \gamma_{q_c}^{q_l}+\lambda^{1}_{lbc}\gamma_{+q_i}^{\phi_l}+\lambda^{2}_{ilc}
\gamma_{+q_b}^{\phi_l}+\lambda^{3}_{ibl}\gamma_{+q_c}^{\phi_l}\right) +      \right. \nonumber\\
& \left. +  \rho_{ibc}\left( \rho_{lbc}\Delta \gamma_{q_j}^{q_l}+\rho_{jlc}
\Delta \gamma_{q_b}^{q_l}+\rho_{jbl}\Delta \gamma_{q_c}^{q_l}+\lambda^{1}_{lbc}\gamma_{+q_j}^{\phi_l}+
\lambda^{2}_{jlc}\gamma_{+q_b}^{\phi_l}+\lambda^{3}_{jbl}\gamma_{+q_c}^{\phi_l}\right)^{*} \right]+\nonumber\\
&-\left(\frac{1}{16\pi^2}\right)^2 \left(2C_a(i)\Delta b_a g_a^4\delta_j^i\right)
\end{align}
where
\begin{equation}
\beta_{g_a}^{\pm}=b_{g_a}^{\pm}\frac{1}{16\pi^2}g_a^3 \qquad \Delta b_a \equiv b_{g_a}^{+}-b_{g_a}^{-}
\end{equation}
and the supersripts $+$ and $-$ refer to the theory above and below the messenger scale respectively.
\begin{align}
\left(\tilde{m}_3^2\right)^{q_i}_{q_j} =\Delta \gamma^{q_i}_{q_k}\gamma_{-q_j}^{q_k}-
\gamma_{-q_k}^{q_i}\Delta \gamma^{q_k}_{q_j}.
\end{align}

\providecommand{\href}[2]{#2}\begingroup\raggedright\endgroup


\begin{thebibliography}{30}

\bibitem{ArkaniHamed:1998kj}
N.~Arkani-Hamed, G.~F. Giudice, M.~A. Luty, and R.~Rattazzi, {\it{Supersymmetry breaking loops from analytic 
continuation into superspace}},{\em Phys.Rev.} {\bf D58} (1998) 115005,[\href{http://xxx.lanl.gov/abs/hep-ph/9803290}
{{\tt hep-ph/9803290}}];
R.~Argurio, M.~Bertolini, G.~Ferretti, and A.~Mariotti, {\it {Patterns of Soft
Masses from General Semi-Direct Gauge Mediation}},  {\em JHEP} {\bf 1003}
(2010) 008, [\href{http://xxx.lanl.gov/abs/0912.0743}{{\tt arXiv:0912.0743}}];
Z.~Komargodski and D.~Shih, {\it {Notes on SUSY and R-Symmetry Breaking in
Wess-Zumino Models}},  {\em JHEP} {\bf 0904} (2009) 093,
[\href{http://xxx.lanl.gov/abs/0902.0030}{{\tt arXiv:0902.0030}}].


\bibitem{Caracciolo:2012de}
F.~Caracciolo and A.~Romanino, {\it{Simple and direct communication of dynamical supersymmetry breaking}}, 
[\href{http://xxx.lanl.gov/abs/1207.5376}{{\tt arXiv:1207.5376}}].



\bibitem{Fayet:1978qc}
P.~Fayet, {\it {Massive gluinos}},  {\em Phys.Lett.} {\bf B78} (1978) 417;
J.~Polchinski and L.~Susskind, {\it {Breaking of Supersymmetry at Intermediate-Energy}},  {\em Phys.Rev.} 
{\bf D26} (1982) 3661;L.~Hall and L.~Randall, {\it {U(1)$_R$ symmetric supersymmetry}},  {\em
Nucl.Phys.} {\bf B352} (1991) 289--308;
P.~J. Fox, A.~E. Nelson, and N.~Weiner, {\it {Dirac gaugino masses and
supersoft supersymmetry breaking}},  {\em JHEP} {\bf 0208} (2002) 035,
[\href{http://xxx.lanl.gov/abs/hep-ph/0206096}{{\tt hep-ph/0206096}}];
H.~Itoyama and N.~Maru, {\it {D-term Dynamical Supersymmetry Breaking Generating Split N=2 Gaugino Masses of 
Mixed Majorana-Dirac Type}}, [\href{http://xxx.lanl.gov/abs/1109.2276}{{\tt arXiv:1109.2276}}]; H.~Itoyama and N.~Maru, 
{\it {D-term Dynamical SUSY Breaking}}, [\href{http://xxx.lanl.gov/abs/1207.7152}{{\tt arXiv:1207.7152}}].










\bibitem{Nardecchia:2009ew}
M.~Nardecchia, A.~Romanino, and R.~Ziegler, {\it {Tree Level Gauge Mediation}},
{\em JHEP} {\bf 0911} (2009) 112,
[\href{http://xxx.lanl.gov/abs/0909.3058}{{\tt arXiv:0909.3058}}];
M.~Nardecchia, A.~Romanino, and R.~Ziegler, {\it {General Aspects of Tree Level
Gauge Mediation}},  {\em JHEP} {\bf 1003} (2010) 024,
[\href{http://xxx.lanl.gov/abs/0912.5482}{{\tt arXiv:0912.5482}}];
M.~Monaco, M.~Nardecchia, A.~Romanino, and R.~Ziegler, {\it {Extended
Tree-Level Gauge Mediation}},  {\em JHEP} {\bf 1110} (2011) 022,
[\href{http://xxx.lanl.gov/abs/1108.1706}{{\tt arXiv:1108.1706}}];
G.~Arcadi, L.~Di Luzio and M.~Nardecchia, {\it {Gravitino Dark Matter in Tree Level Gauge Mediation with and 
without R-parity}}, {\em JHEP} {\bf 1112} (2011) 040, [\href{http://xxx.lanl.gov/abs/1110.2759} {{\tt arXiv:1110.2759}}].


\bibitem{Seiberg:2008qj}
N.~Seiberg, T.~Volansky, and B.~Wecht, {\it {Semi-direct Gauge Mediation}},
{\em JHEP} {\bf 0811} (2008) 004, [\href{http://xxx.lanl.gov/abs/0809.4437}{{\tt arXiv:0809.4437}}];
H.~Elvang and B.~Wecht, {\it {Semi-Direct Gauge Mediation with the 4-1 Model}}, {\em JHEP} {\bf 0906} (2009) 026, 
[\href{http://xxx.lanl.gov/abs/0904.4431}{{\tt arXiv:0904.4431}}].



\bibitem{Dine:1995ag}
M.~Dine, A.~E.~Nelson, Y.~Nir and Y.~Shirman, {\it{New tools for low-energy dynamical supersymmetry breaking}}, 
{\em Phys.Rev.}  {\bf D53} (1996) 2658, [\href{http://xxx.lanl.gov/abs/hep-ph/9507378}{{\tt hep-ph/9507378}}];
E.~Poppitz and S.~P.~Trivedi, {\it{Some examples of chiral moduli spaces and dynamical supersymmetry breaking}}, 
{\em Phys.Lett.} {\bf B365} (1996) 125, [\href{http://xxx.lanl.gov/abs/hep-th/9507169}{{\tt hep-th/9507169}}].


\bibitem{Craig:2012xp}
N.~Craig, S.~Knapen, D.~Shih and Y.~Zhao, 
{\it{A Complete Model of Low-Scale Gauge Mediation}}, [\href{http://xCraig:2012xpxx.lanl.gov/abs/1206.4086}{{\tt arXiv:1206.4086}}];
A.~Albaid and K.~S.~Babu, 
{\it{Higgs boson of mass 125 GeV in GMSB models with messenger-matter mixing}}, [\href{http://xxx.lanl.gov/abs/arXiv:1207.1014}{{\tt arXiv:1207.1014}}].





\bibitem{Chacko:2001km}
Z.~Chacko and E.~Ponton, {\it{Yukawa deflected gauge mediation}}, {\em Phys.Rev.}  {\bf D66} (2002) 095004,
 [\href{http://xxx.lanl.gov/abs/hep-ph/0112190}{{\tt hep-ph/0112190}}]; M.~Abdullah, I.~Galon, Y.~Shadmi and Y.~Shirman,
  {\it{Flavored Gauge Mediation, A Heavy Higgs, and Supersymmetric Alignment}}, [\href{http://xxx.lanl.gov/abs/1209.4904}{{\tt arXiv:1209.4904}}].













\bibitem{Dine:1981za}
M.~Dine, W.~Fischler, and M.~Srednicki, {\it {Supersymmetric Technicolor}},
{\em Nucl.Phys.} {\bf B189} (1981) 575--593;
M.~Dine and W.~Fischler, {\it {A Phenomenological Model of Particle Physics
Based on Supersymmetry}},  {\em Phys.Lett.} {\bf B110} (1982) 227;
M.~Dine and A.~E. Nelson, {\it {Dynamical supersymmetry breaking at
low-energies}},  {\em Phys.Rev.} {\bf D48} (1993) 1277--1287,
[\href{http://xxx.lanl.gov/abs/hep-ph/9303230}{{\tt hep-ph/9303230}}];
M.~Dine, A.~E. Nelson, and Y.~Shirman, {\it {Low-energy dynamical supersymmetry
breaking simplified}},  {\em Phys.Rev.} {\bf D51} (1995) 1362--1370,
[\href{http://xxx.lanl.gov/abs/hep-ph/9408384}{{\tt hep-ph/9408384}}].





\bibitem{Banks:2010zn}
T.~Banks and N.~Seiberg, {\it{Symmetries and Strings in Field Theory and Gravity}}, {\em Phys.Rev.}  {\bf D83} (2011) 084019,
[\href{http://xxx.lanl.gov/abs/1011.5120}{{\tt arXiv:1011.5120}}].
; the novelty of th

\bibitem{Giudice:1998bp}
G.~F.~Giudice and R.~Rattazzi, {\it{Theories with gauge mediated supersymmetry breaking}}, 
{\em Phys.Rept.}  {\bf 322} (1999) 419, [\href{http://xxx.lanl.gov/abs/hep-ph/9801271}{{\tt hep-ph/9801271}}].



\bibitem{Brizi:2009nn} 
L.~Brizi, M.~Gomez-Reino and C.~A.~Scrucca, {\it{Globally and locally supersymmetric effective theories for light fields}}, {\em Nucl.Phys.} {\bf B820}, 193 (2009), [\href{http://xxx.lanl.gov/abs/arXiv:0904.0370}{{\tt arXiv:0904.0370}}].










\end{thebibliography}
\end{document}